\documentstyle[12pt,aaspp]{article}

\begin{document}

%
%
%
   \def\prd#1#2#3#4{#4 19#3 Phys.~Rev.~D,\/ #1, #2 }
   \def\pl#1#2#3#4{#4 19#3 Phys.~Lett.,\/ #1, #2 }
   \def\prl#1#2#3#4{#4 19#3 Phys.~Rev.~Lett.,\/ #1, #2 }
   \def\pr#1#2#3#4{#4 19#3 Phys.~Rev.,\/ #1, #2 }
   \def\prep#1#2#3#4{#4 19#3 Phys.~Rep.,\/ #1, #2 }
   \def\pfl#1#2#3#4{#4 19#3 Phys.~Fluids,\/ #1, #2 }
   \def\pps#1#2#3#4{#4 19#3 Proc.~Phys.~Soc.,\/ #1, #2 }
   \def\nucl#1#2#3#4{#4 19#3 Nucl.~Phys.,\/ #1, #2 }
   \def\mpl#1#2#3#4{#4 19#3 Mod.~Phys.~Lett.,\/ #1, #2 }
   \def\apj#1#2#3#4{#4 19#3 Ap.~J.,\/ #1, #2 }
   \def\aj#1#2#3#4{#4 19#3 Astr.~J.,\/ #1, #2}
   \def\acta#1#2#3#4{#4 19#3 Acta ~Astr.,\/ #1, #2}
   \def\rev#1#2#3#4{#4 19#3 Rev.~Mod.~Phys.,\/ #1, #2 }
   \def\nuovo#1#2#3#4{#4 19#3 Nuovo~Cimento~C,\/ #1, #2 }
   \def\jetp#1#2#3#4{#4 19#3 Sov.~Phys.~JETP,\/ #1, #2 }
   \def\sovast#1#2#3#4{#4 19#3 Sov.~Ast.~AJ,\/ #1, #2 }
   \def\pasj#1#2#3#4{#4 19#3 Pub.~Ast.~Soc.~Japan,\/ #1, #2 }
   \def\pasp#1#2#3#4{#4 19#3 Pub.~Ast.~Soc.~Pacific,\/ #1, #2 }
   \def\annphy#1#2#3#4{#4 19#3 Ann. Phys. (NY), \/ #1, #2 }
   \def\yad#1#2#3#4{#4 19#3 Yad. Fiz.,\/ #1, #2 }
   \def\sjnp#1#2#3#4{#4 19#3 Sov. J. Nucl. Phys.,\/ #1, #2 }
   \def\astap#1#2#3#4{#4 19#3 Ast. Ap.,\/ #1, #2 }
   \def\anrevaa#1#2#3#4{#4 19#3 Ann. Rev. Astr. Ap.,\/ #1, #2
                       }
   \def\mnras#1#2#3#4{#4 19#3 M.N.R.A.S.,\/ #1, #2 }
   \def\jdphysics#1#2#3#4{#4 19#3 J. de Physique,\/ #1,#2 }
   \def\jqsrt#1#2#3#4{#4 19#3 J. Quant. Spec. Rad. Transfer,\/ #1,#2 }
   \def\jetpl#1#2#3#4{#4 19#3 J.E.T.P. Lett.,\/ #1,#2 }
   \def\apjl#1#2#3#4{#4 19#3 Ap. J. (Letters).,\/ #1,#2 }
   \def\apjs#1#2#3#4{#4 19#3 Ap. J. (Supp.).,\/ #1,#2 }
   \def\apl#1#2#3#4{#4 19#3 Ap. Lett.,\/ #1,#2 }
   \def\astss#1#2#3#4{#4 19#3 Ap. Sp. Sci.,\/ #1,#2 }
   \def\nature#1#2#3#4{#4 19#3 Nature,\/ #1,#2 }
   \def\spscirev#1#2#3#4{#4 19#3 Sp. Sci. Rev.,\/ #1,#2 }
   \def\advspres#1#2#3#4{#4 19#3 Adv. Sp. Res.,\/ #1,#2 }
   %
%
%
\def\Msun{M_{\odot}}
\def\Mdot{\dot M}
\def\deg{$^\circ$\ }
\def\etal{{\it et~al.\ }}
\def\eg{{\it e.g.,\ }}
\def\etc{{\it etc.}}
\def\ie{{\it i.e.,}\ }
\def\ksec{{km~s$^{-1}$}}
\def\arcsec{{$^{\prime\prime}$}}
\def\arcmin{{$^{\prime}$}}
\def\subsun{_{\twelvesy\odot}}
\def\sun{\twelvesy\odot}
\def\gtwid{\mathrel{\raise.3ex\hbox{$>$\kern-.75em\lower1ex\hbox{$\sim$}}}}
\def\ltwid{\mathrel{\raise.3ex\hbox{$<$\kern-.75em\lower1ex\hbox{$\sim$}}}}
\def\plusminus{\mathrel{\raise.3ex\hbox{$+$\kern-.75em\lower1ex\hbox{$-$}}}}
\def\minusplus{\mathrel{\raise.3ex\hbox{$-$\kern-.75em\lower1ex\hbox{$+$}}}}

\title{The Vertical Structure and Ultraviolet Spectrum of Accretion Disks
Heated by Internal Dissipation in Active Galactic Nuclei}

\author{Mark W. Sincell}
\affil{Department of Physics MC 704 \\The University of Illinois at
Urbana-Champaign \\1110 W. Green Street \\Urbana, IL 61801-3080}
\authoraddr{Department of Physics MC 704 \\The University of Illinois at
Urbana-Champaign \\1110 W. Green Street \\Urbana, IL 61801-3080}

\author{Julian H. Krolik}
\affil{Department of Physics \& Astronomy \\The Johns Hopkins University 
\\Baltimore, MD 21218}
\authoraddr{Department of Physics \& Astronomy \\The Johns Hopkins University 
\\Baltimore, MD 21218}

\begin{abstract} 

We present an improved calculation of the vertical structure and
ultraviolet spectrum of a dissipative accretion disk in an AGN. We
calculate model spectra in which the viscous stress is proportional to the
total pressure, the gas pressure only and the geometric mean of the
radiation and gas pressures (cf. Laor \& Netzer 1989: LN89).  As a result
of a more complete treatment of absorptive opacity, we find greater
overall spectral curvature than did LN89, as well as larger amplitudes in
both the Lyman and HeII photoionization edges.  The local black body
approximation is not a good description of the near UV spectrum.  With
relativistic corrections (appropriate to non-rotating black holes)
included, we find that the near UV spectrum hardens with increasing $\dot
m / m_8$ ($\dot m$ is the accretion rate in Eddington units, $m_8$ the
black hole mass in units of $10^8 M_{\odot}$).  The near UV spectrum is
consistent with observations if $\dot m m_8^{-1} \sim 10^{-3}$, but disks
this cold would have large, and unobserved, absorption features at the
Lyman edge.  The edge amplitude is reduced when $\dot m/m_8$ is larger,
but then the near-UV slope is too hard to match observations.  We conclude
that models in which conventional disks orbit non-rotating black holes do
not adequately explain UV continuum production in AGN. 

\end{abstract}

\section{Introduction}
\label{sec: introduction}

The optical and UV emission of radio-quiet active galactic nuclei (AGNs) 
is dominated by a quasi-thermal component, the ``Big Blue Bump" (Shields
1978, Malkan \& Sargent 1982, Malkan 1983).  The Big Blue Bump is usually
interpreted as thermal emission from a geometrically thin, optically
thick, accretion disk around a massive black hole (Malkan 1983, Sun \&
Malkan 1989, Laor \& Netzer 1989: LN89).  There are two general reasons
for this belief.  First, if the accreting gas has any angular momentum far
from the black hole, it will collapse to a disk as it falls into the
gravitational potential of the hole.  Second, the effective temperature of
radiation from optically thick gas accreting onto a $10^8$ solar mass
black hole at the Eddington accretion rate is roughly $T_{eff} \sim
10^5$~K.  Black body emission at $10^5$~K peaks in the optical and UV
bands.  Unfortunately, theoretical models of accretion disk spectra do not
agree with the observed optical-UV spectra of AGNs. 

A benchmark calculation of the vertical structure and optical-UV spectrum
of an accretion disk around a massive black hole was performed by LN89
(see also Laor, Netzer \& Piran 1990 and Laor 1990).  They calculated the
radial structure of the disk for three different viscosity prescriptions
and computed the corresponding emergent spectra with a semi-analytic
approximation to the radiative transfer equation. The predicted slope of
the near UV spectrum ($\alpha_n \sim 0.1$, where $F(\nu) \propto
\nu^{\alpha_n}$) was larger than the mean near UV spectral index for a
sample of $\sim 100$ AGNs ($\alpha_{obs} \sim -0.5$, Laor 1990).  LN89
also predicted that Lyman edge features, both in absorption and emission,
should generally be present in accretion disk spectra.  However, Lyman
edge features have only been detected in a few AGNs (Antonucci, Kinney \&
Ford 1989, Koratkar, Kinney \& Bohlin 1992). 

Since the work of LN89, the sophistication of numerical models of AGN
accretion disks has increased dramatically, without a corresponding
improvement in fits to the observed UV spectra.  These models can be
broadly divided into two groups: those which compute the UV spectrum using
the diffusion approximation (Ross, Fabian \& Mineshige 1992, 
Shimura \& Takahara 1993, D\"orrer, \etal 1996) and those which explicitly
solve the radiative
transfer equations (Hubeny \& Hubeny 1997, Blaes \& Agol 1996, Storzer,
Hauschildt \& Allard 1994).  

In this paper, we present a calculation of the UV spectra of accretion
disks around massive non-rotating black holes using the method developed
by Sincell \& Krolik (1997).  We solve the frequency- and angle-dependent
radiative transfer equation with a variable Eddington factor method (Auer
\& Mihalas 1970).  We assume LTE level populations and neglect
Comptonization. The gas temperature is determined by balancing viscous and
radiative heating with radiative cooling and the density is determined by
hydrostatic equilibrium. The local disk spectra are then added together,
including Doppler boosting by the rotation of the disk and the redshift
caused by propagation out of the potential well of the black hole, to form
the observed integrated disk spectrum.

\section{Method}
\label{sec: method}

The emergent flux from one face of a geometrically thin
Keplerian disk around a black hole is
\begin{equation}
\label{eq: dissipation rate}
Q_{dis} = {3GM \dot M \over 8\pi R^3}
\left( {Q_{NT} \over B_{NT} C_{NT}^{1/2} } \right),
\end{equation}
independent of the viscosity.  The relativistic correction factors
$Q_{NT}$, $B_{NT}$, and $C_{NT}$ are taken
from the standard references (Novikov \& Thorne 1973, Page \& Thorne
1974), and apply to accretion disks around non-rotating (Schwarzschild)
black holes.

The surface density required to support a given accretion rate
is determined by relating the vertically-averaged
$r$-$\phi$ component of the stress tensor to local physical conditions.
The conventional {\it ansatz} for doing this is the ``$\alpha$"-model
introduced by Shakura \& Sunyaev (1973), in which this component of
the stress is taken to be
a dimensionless number $\alpha_{SS}$ times the vertically-averaged pressure.
However, as Shakura \& Sunyaev (1976) showed, this model is thermally
unstable when, as commonly occurs, the disk pressure is dominated by
radiation.  

In the absence of a compelling physical argument for any particular
viscosity prescription, we follow LN89 and investigate three different
stress prescriptions.  We calculate the disk surface mass density assuming
that the stress tensor is proportional to the total (gas plus radiation)
pressure, the gas pressure and the geometric mean of the radiation and
gas pressures.  The second and third models are thermally stable.  We
choose
$\alpha_{SS} = 0.1$,
but the results depend only very weakly on the value of this
number.  

Under these assumptions, the (half) surface mass density at
a given radius $r = R/R_s$, where $R_s = 2GM/c^2$ is the
Schwarzschild radius, is 
\begin{equation}
\label{eq: sigma pt}
\Sigma_T = 0.723 \alpha_{SS}^{-1} \dot m^{-1} r^{3/2}
\left( { B_{NT}^3 C_{NT}^{1/2} E_{NT} \over A_{NT}^{2}
Q_{NT} } \right)
\mbox{gm ${\rm cm^{-2}}$}.
\end{equation}
for $t_{r \phi} \sim P_r + P_g$ (the $P_t$ case),
\begin{equation}
\label{eq: sigma pm}
\Sigma_T = 5.63 \times 10^3 \alpha_{SS}^{-8/9} \dot m^{1/9} m_8^{-1/9}
r^{1/3}
\left( { B_{NT} C_{NT}^{1/2} E_{NT}^{4/9} \over A_{NT}^{8/9} Q_{NT}^{1/9} }
\right)
\mbox{gm ${\rm cm^{-2}}$}
\end{equation}
for $t_{r\phi} \sim (P_r P_g)^{1/2}$ (the $P_m$ case) and
\begin{equation}
\label{eq: sigma pg}
\Sigma_T = 1.27 \times 10^{7} \alpha_{SS}^{-4/5} \dot m^{3/5} m_8^{1/5}
r^{-3/5} 
\left( { C_{NT}^{1/2} Q_{NT}^{3/5} \over B_{NT}^{3/5} D_{NT}^{4/5} }
\right)
\mbox{gm ${\rm cm^{-2}}$}
\end{equation}
for $t_{r\phi} \sim P_g$ (the $P_g$ case).
In these expressions we introduce $\dot m$, the accretion rate
in units of the Eddington critical accretion rate assuming the Schwarzschild
efficiency of 0.057, and $m_8$,
the mass of the central black hole in units of $10^8$ solar masses
(\eg SK97).  The relativistic correction factors
are taken from Novikov \& Thorne (1973) and Riffert \& Herold (1995).

To find the local heating rate per unit mass, we assume that at any
given radius the heating rate is independent of position.  It is then
\begin{equation}
\label{eq: local heating rate}
{d Q \over d \Sigma} = {2 Q_{dis} \over \Sigma_T},
\end{equation}
where the factor of two reflects the fact that $\Sigma_T$ is the total
surface mass density of the disk and we are calculating the flux from
only one face of the disk.

We determine the vertical structure and UV spectrum of the disk by solving
the coupled differential equations describing the transfer of radiation,
hydrostatic equilibrium, and the relation between optical depth
and the vertical coordinate, subject to the local constraints
of thermal equilibrium and charge conservation.  We include electron
scattering, bremsstrahlung and HI, HeI and HeII photoionization opacities 
and the corresponding continuum cooling processes.  We neglect all line
opacities and emissivities. A detailed description of
the computational method is presented in SK97.  

In order to calculate the spectrum of a conventional disk, we have made
two changes to the method described in SK97. First, the local heating rate
is determined by viscous heating (eq. \ref{eq: local heating rate}).  In
the present paper, we neglect irradiation of the disk.  Second, the
boundary conditions on the radiation flux and the gas pressure have been
changed.  The gas pressure at the upper edge of the disk is set to a small
fraction of the total central pressure.  The radiation flux is set to zero
at the disk midplane. 

\section{Results}
\label{sec: results}

\subsection{Vertical Structure of the Accretion Disk}
\label{subsec: vertical structure}

The analytic solution used by LN89 for the gas density (their equations 13,
along with their assumption of constant density as a function of altitude)
and temperature (as expressed in their equation 25) is
in good agreement with our calculation because electron scattering
dominates the flux weighted mean opacity.  A typical example of the
results of a vertical structure calculation is shown in fig. \ref{fig: tp
profile}.  The dashed lines are the gas density and temperature profiles
used by LN89.  The gas density is constant near the midplane of the
disk, as expected for a radiation pressure dominated disk, but
drops rapidly above $\Sigma \sim 100\mbox{gm ${\rm cm^{-2}}$}$.  The LN89
estimate of the gas density at the midplane is about a factor of two
smaller than the actual central density and about an order of magnitude
larger than the density at $\Sigma = 1 \mbox{gm ${\rm cm^{-2}}$}$.  The
discrepancy at the midplane is due to the Riffert \& Herrold (1995)
corrections to the expression for the vertical gravity in the relativistic
portion of a thin disk, which were not included in LN89. 
The numerical value for the midplane density agrees with the LN89
value to within a few percent when we use the same expression for
the vertical gravity as they used.  The gas
temperature estimate of LN89 is a good approximation to the true profile
at large optical depths, but the more accurate treatment gives
temperatures which are about $10 \%$ higher near the top of the disk.

\subsection{Accretion Disk Spectra}
\label{subsec: spectrum}
 
Despite the close agreement between the gas pressure and temperature
profiles, our predicted spectrum is quite different from the one found
by LN89.  The reason for the discrepancy is that the
LN89 calculation of the emergent spectrum is accurate only if electron 
scattering dominates the opacity at {\it all} frequencies.
They define the mass column $\Sigma_{ph}$ of the UV photosphere at $\nu$ by
setting the effective optical depth of the photosphere
\begin{equation}
\tau_{ph} = \left( {\kappa_{a,\nu} \over \kappa_{a,\nu} + \kappa_{es} } 
         \right)^{1/2} \kappa_{es} \Sigma_{ph}
       = {2 \over 3}.
\end{equation}
When $\kappa_{a,\nu} \gtwid \kappa_{es}$, this expression gives the
erroneous result that $\tau_{ph} = \kappa_{es} \Sigma_{ph}$, rather than
the correct optical depth, $\tau_{ph} = \kappa_{a,\nu} \Sigma_{ph}$.
As a result, this prescription overestimates $\Sigma_{ph}$ by a factor
$\kappa_{a,\nu}/\kappa_{es}$.

Because the temperature increases with increasing $\Sigma$, this method
overestimates the photospheric temperature, and the observed flux at $\nu$, 
when $\kappa_{a,\nu} \gtwid \kappa_{es}$.
We find that the absorptive opacity dominates several significant frequency
ranges---well below the Lyman edge (where free-free opacity is important)
and just
above the HeII edge (HeII photoionization opacity), even though electron
scattering dominates the flux-weighted opacity.  Therefore, the largest
differences between our calculation and LN89 occur in these frequency ranges.  

Our improvements in the radiative transfer calculation result in three
changes to the predicted spectrum of an accretion disk (fig.  \ref{fig:
laor comparison}).  First, we find that the slope of the non-ionizing
continuum is much harder than predicted by LN89.  The free-free opacity
$\kappa_{ff} \gtwid \kappa_{es}$ well below the Lyman edge, so the mass
column of the photosphere increases with frequency ($\kappa_{ff} \propto
\nu^{-3}$, Rybicki \& Lightman 1979). This results in a rapid increase in
the flux density with frequency because radiation from higher frequencies
comes from hotter gas.  In contrast, the LN89 prescription results in a
nearly constant $\Sigma_p$ in this frequency range and a much weaker
frequency dependence for the flux.  Second, the amplitudes of the
spectral features are underestimated by the LN89 model.  The opacity jump
across the Lyman edge is large in both models but, for the reasons
outlined above, LN89 underestimate the corresponding change in
photospheric temperature because $\kappa_{a,\nu} \gtwid \kappa_{es}$. 
Third, the continuum flux above the HeII edge is overestimated by LN89. 
The HeII photoionization opacity is much larger than $\kappa_{es}$ at $r
\ltwid 10$, where most of the high frequency flux is formed, which causes
the LN89 approach to overestimate the photospheric temperature. 
 
The surface mass density of the disk is smallest for the $P_t$ case and
largest for the $P_g$ case (eqs. \ref{eq: sigma pt} - \ref{eq: sigma pg}).
Although the difference in $\Sigma_T$ can be many orders of
magnitude, we find that the spectral changes are comparatively small
(fig. \ref{fig: standard visc}).  The gas density increases with $\Sigma$,
so the bound-free opacity of the $P_t$ case is
much smaller than either the $P_{m}$ or $P_g$ cases.
Compared to the other cases, the flux above the HeII edge is about a
factor of ten larger in the $P_t$ case, with a
corresponding decrease of a factor of two in flux below the Lyman edge.
The flux between
these frequencies is nearly the same in all three cases.  The shift of
flux from low to high frequencies results in smaller edge features for the
$P_t$ case.

In figures \ref{fig: standard mass} and \ref{fig: standard mdot} we plot the
spectra of the standard accretion disk as a function of $m_8$ for fixed 
$\dot m$ and as a function of $\dot m$ for fixed $m_8$, respectively.
In order to illustrate the effects of variations in these parameters, we
fix the viscosity prescription and use the thermally stable $P_m$ case.
Although there is continuous curvature to the spectra, they can
be characterized by three parameters: the near ($\alpha_n$) and far
($\alpha_f$) UV spectral
indices and the amplitude of the Lyman edge feature ($A_l$). We define the
spectral indices as 
\begin{equation}
\alpha_{n,f} = {\log F(\nu_2) - \log F(\nu_1) \over \log \nu_2 -
\log \nu_1},
\end{equation}
where $\log\nu_1 = 14.8$ and $\log\nu_2 = 15.5$ for $\alpha_n$, and
$\log\nu_1 = 15.5$ and $\log\nu_2 = 16.0$ for $\alpha_f$.   The amplitude
of the Lyman edge feature is defined as the ratio of the fluxes just above
and below the edge.  This quantity is less than one for absorption
edges and larger than one for emission edges.

The shape of the spectrum depends primarily on the ratio of the accretion
rate to the central mass, $\dot m / m_8$ (see table \ref{table:
spectral parameters}). This quantity determines 
the effective temperature scale of the disk because  $T_{eff} \propto
Q_{dis}^{1/4} \propto (\dot m / m_8)^{3/4}$.  We find that the near UV
spectral index hardens from $\alpha_n \simeq -0.6$ for $\dot m / m_8 =
0.001$ and $m_8 = 27.0$ to $\alpha_n \simeq 0.4$ for $\dot m / m_8 = 0.1$ 
and $m_8 = 0.27$. For the same
range of $\dot m / m_8$, the far UV spectral index increases from
$\alpha_f \sim - \infty$ (no flux above the Lyman edge) to $\alpha_f
\simeq -1.2$ and the amplitude of the Lyman edge increases from $A_l \sim 
0$ (no flux above the Lyman edge) to $A_l \simeq 1.6$.  The near UV
spectral index is also weakly dependent on the disk luminosity ($\propto
\dot m m_8$) so that the spectrum softens slightly with increasing disk
luminosity. 
No appreciable emission is seen above the HeII edge in any of the disks.
Only in the hottest case ($\dot m = 0.3, m_8 = 2.7$) is any emission seen
at those frequencies.

Although the near UV spectral index of the hotter disks is near the
canonical $\alpha_n = 1/3$ power law, appropriate for a Newtonian disk
(\eg Pringle 1981), this is a coincidence caused by two competing effects.
On the one hand, we find that the near-UV spectrum of an individual ring
is significantly harder than a black body at its effective temperature.
On the other hand (as pointed out
by LN89), the general relativistic corrections to the disk
structure equations soften the disk-integrated spectrum by making the
dependence of temperature on radius shallower.

\subsection{Check of the Model Assumptions}
\label{subsec: assumption check}

The computed gas temperature and pressure profiles also allow us to make
an {\it a posteriori} check of our model assumptions: LTE level
populations and no Comptonization.

Comptonization will change the observed disk spectrum if
the Compton opacity is larger than the absorptive opacity and the Compton
y-parameter, defined as $y \equiv 4kT/(m_e c^2)\max(\tau_T,\tau_{T}^2)$
(Rybicki and Lightman 1979), is greater than unity.  In the present
context, the relevant Compton optical depth is the optical
depth from the photosphere to the disk surface, so
$\tau_T = \Sigma_{ph}\kappa_{es}$.   For most frequencies
$\tau_T \gg 1$.  In the spectral range where free-free opacity is the
dominant absorption mechanism, yet $\kappa_{ff} < \kappa_{es}$, we
define $y_{ff}$ by
\begin{equation}
  y_{ff} = 0.055 h_{13}^{2/3} T_{5}^{10/3} {\omega^2 \over (
1-e^{-\omega})^{2/3}}
\end{equation}
where $\omega = h\nu / kT$, $h_{13}$ is the scale height for changes is
the gas density near the photosphere in
$10^{13}$~cm and $T_5$ is the disk temperature in $10^5$~K.  

For our range of central masses and accretion rates, the height of the
radiation pressure supported disk is $\ltwid 10^{14}$~cm. At $\Sigma
\gtwid 100$~gm/cm$^2$ gas pressure gradients support the disk against
gravity and we find that the scale height is typically $10^{-3}$ times the
thickness of the disk.  The photosphere lies in the gas pressure supported
layer of the disk for virtually all frequencies in all our models,
implying that $h_{13} \ltwid 10^{-2}$. The photospheric temperature is $T
\ltwid 10^5$~K, and we conclude that $y_{ff} < 10^{-3}$ for all our
models.  Including bound-free opacity in this estimate would further
reduce $y_{ff}$.  Note, however, the very sharp dependence of $y$
on temperature.  Smaller central masses, which lead to higher temperatures,
can create circumstances in which Comptonization is significant (e.g. as
in the work of Ross \etal 1992).

Departures from LTE level populations become significant when the gas
pressure in the disk is $P \ltwid 100$~dyn~cm$^{-2}$
(Storzer, \etal 1994).  When
the gas pressure is this low, non-LTE effects can reduce the amplitude of
the Lyman edge or drive it into emission.  However, we expect LTE to be a
good approximation in our case because $P \gtwid 100$~dyn~cm$^{-2}$ near the
photospheres of our models.  To check this assumption, we have made a
non-LTE calculation of the disk spectrum for the case $m_8=2.7$ and $\dot
m = 0.3$.  As expected, we find that non-LTE effects have a negligible
effect upon the spectrum.

\section{Conclusion}
\label{sec: conclusion}

We have calculated the vertical structure and UV spectrum of a
conventional, viscously heated, accretion disk around a massive
non-rotating black
hole.  Electron scattering dominates the opacity and the disk
is supported by radiation pressure, except in the uppermost layers, where
the gravitational acceleration becomes large and gas pressure gradients
support the disk.  The computed midplane density and temperature agree
with the predictions of analytic calculations of the vertical structure
(LN89, Shakura \& Sunyaev 1973), if the Riffert \& Herrold (1995)
corrections to the vertical gravity are included. 

Although the flux weighted mean opacity is dominated by electron
scattering, the absorptive opacity can exceed the Thomson opacity both
below the Lyman edge and above the HeII edge.  In contrast, the best
previous calculation (LN89) assumed that electron scattering dominates the
opacity at all frequencies, leading them to overestimate the flux in these
frequency ranges. As a consequence of our improved treatment of the gas
opacity and radiative transfer, we find greater overall spectral curvature
and larger features at the Lyman edge and the HeII edge.  Although the
spectral index of the near UV spectrum of the hotter disks is near the
Newtonian black body value ($\alpha_n = 1/3$), this is just a coincidence. 
We find that the local black body approximation is not a good description
of the disk spectrum. 

We computed spectra assuming that the viscous stress was proportional to
the total (gas plus radiation) pressure, the gas pressure and the
geometric mean of the two pressures.  We find that the lower densities in
the $P_t$ case cause the flux above the HeII edge to increase by about 
an order of magnitude and the flux below the Lyman edge to decrease by
about a factor of two. This results in somewhat
smaller edge amplitudes in this case.  The flux between these two
energies is unaffected by the viscosity prescription.

The shape of the UV spectrum is determined by the effective temperature of
the disk radiation or, equivalently, $\dot m m_8^{-1}$.  For the hottest
disks, the near UV spectral index, $\alpha_n \simeq 0.4$ and the far UV
index, $\alpha_f \simeq -1.4$.  The Lyman edge can be in emission for
these disks.  The coldest disks have softer near UV spectra ($\alpha_n
\simeq -0.6$) and are completely absorbed above the Lyman edge.

The average optical-UV power-law index for quasars is $\sim -0.5$ (Laor
1990)  but a single power law is a poor description of the continuum
spectrum (Francis, \etal 1991).  The continuum of the composite quasar
spectrum, formed by combining all the quasars in the Large Bright Quasar
Survey, softens with increasing frequency (Francis, \etal 1991).  Both the
mean optical-UV slope and the spectral curvature are consistent with our
predicted UV spectrum of an irradiated disk if $\dot m m_8^{-1} \sim
10^{-3}$.  However, disks this cold would be completely absorbed above
the Lyman edge, which is not observed (Antonucci, Kinney
\& Ford 1989; Koratkar, Kinney \& Bohlin 1992).  On the other hand,
disks hot enough to eliminate the Lyman edge feature all have $\alpha_n$
much larger than the observed slope.

  Observational support for the existence of accretion disks in AGN
originally came from the qualitative match between the ``Big Blue Bump"
and the characteristic temperature scale predicted by the thermal accretion
disk model.  We now see that the simplest theoretical models fail when
confronted with more specific spectral tests.  The model considered here,
in which the dissipation all occurs inside a geometrically thin, but
optically very thick, disk around a non-rotating black hole
does not adequately explain the observed UV
spectrum.  

The next step in this program is to study the consequences of placing
accretion disks around rotating (Kerr) black holes.  Laor (1992)
argued that the large Doppler shifts and gravitational
redshifts of disk radiation
from near the last stable orbit of a Kerr black hole could spread the
Lyman edge feature and make it undetectable, but did not present detailed
calculations of these effects.  In addition, the effective
temperature distribution for a disk around a Kerr black hole is different
from that for a disk around a Schwarzschild hole, and this, too,
may affect the  shape of the UV continuum.   We plan to construct models
for the structure and UV continuum emission from disks around Kerr black
holes in the near future.

\acknowledgements 

We would like to thank Ari Laor for helpful conversations and for
providing many disk spectra.  We also thank Omer Blaes and Eric Agol for
ongoing conversations.  MWS received support for this work from NASA
grants NAGW-3129, 1583 and NAG 5-2925, and NSF grant AST 93-15133.  JHK
was partially supported by NASA Grant NAGW-3156.

\vfil
\eject
\centerline{Figure and Table Captions}

\begin{figure}
\plotone{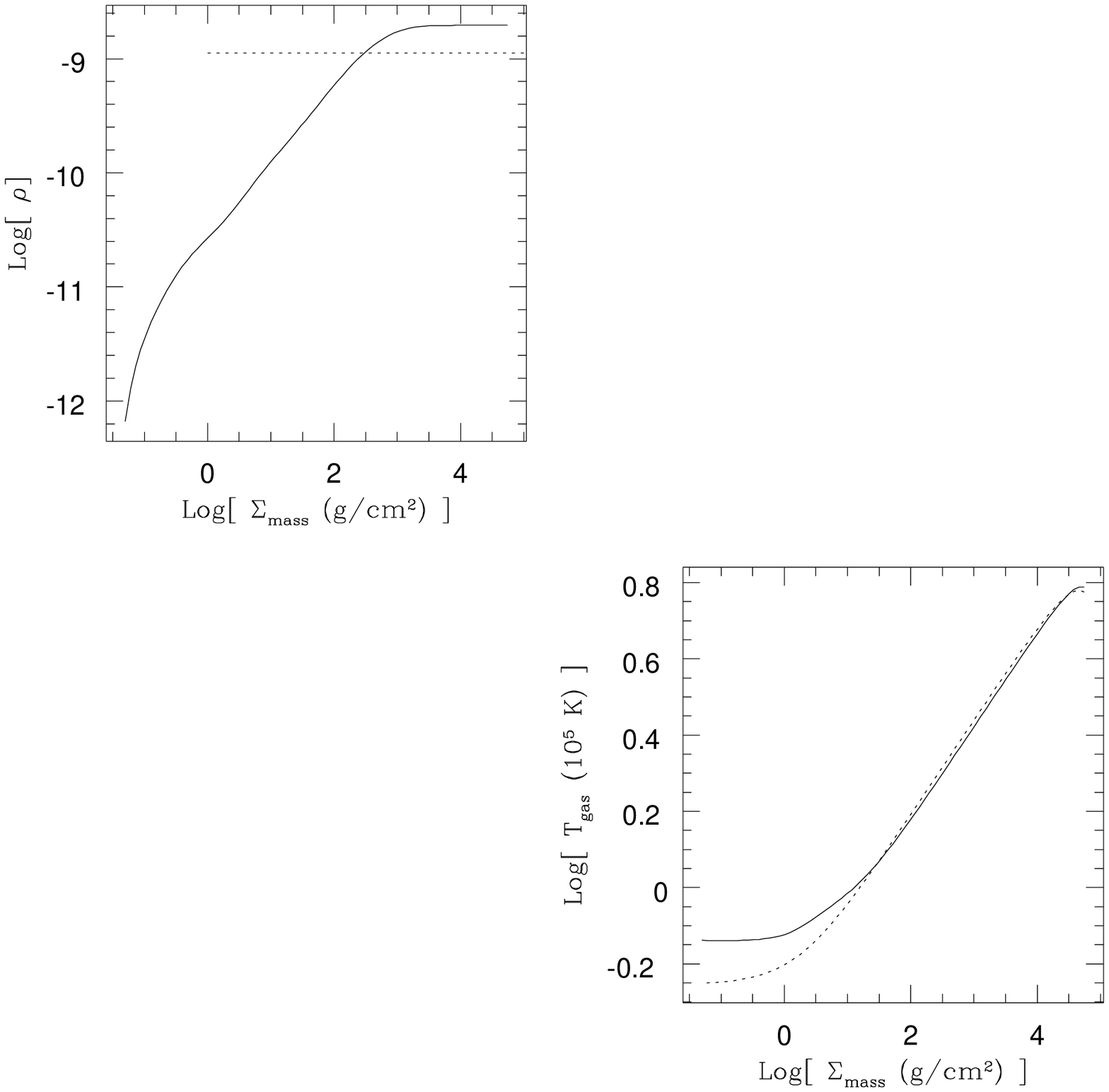}
\caption{\label{fig: tp profile} }
\end{figure}

\bigskip
Figure \ref{fig: tp profile}.  The density (a) and temperature (b) profiles
for a standard accretion disk.  Model parameters are $\dot m=0.3$, $m_8 
= 2.7$ and $r=6.8$.  The dashed line is the analytic model of LN89.

\begin{figure}
\plotone{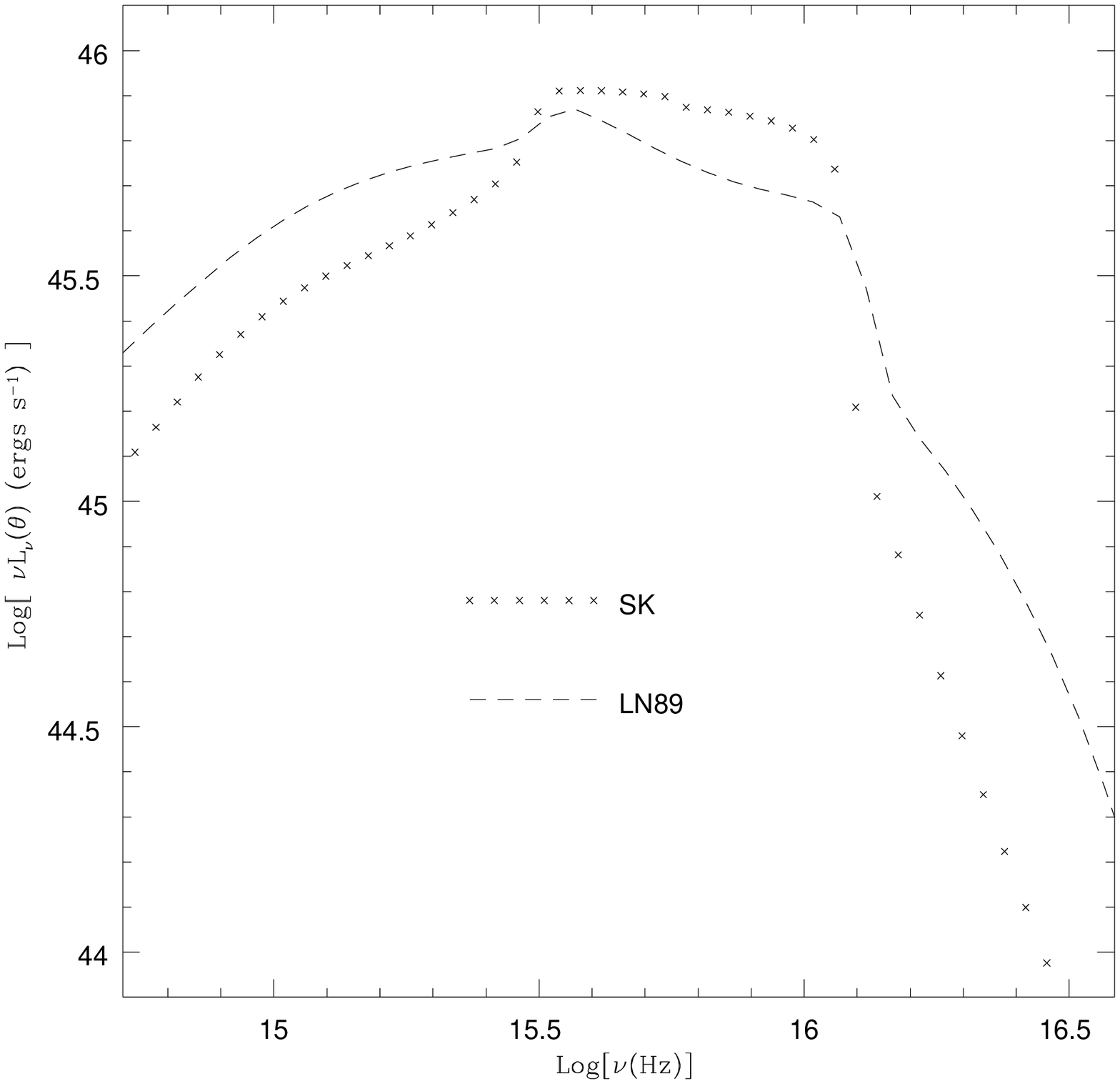}
\caption{\label{fig: laor comparison} }
\end{figure}

\bigskip
Figure \ref{fig: laor comparison}.  Comparison of the face-on disk spectra 
calculated by LN89 and in this paper.
The disk parameters are $m_8 = 2.7$ and $\dot m = 0.3$.  

\begin{figure}
\plotone{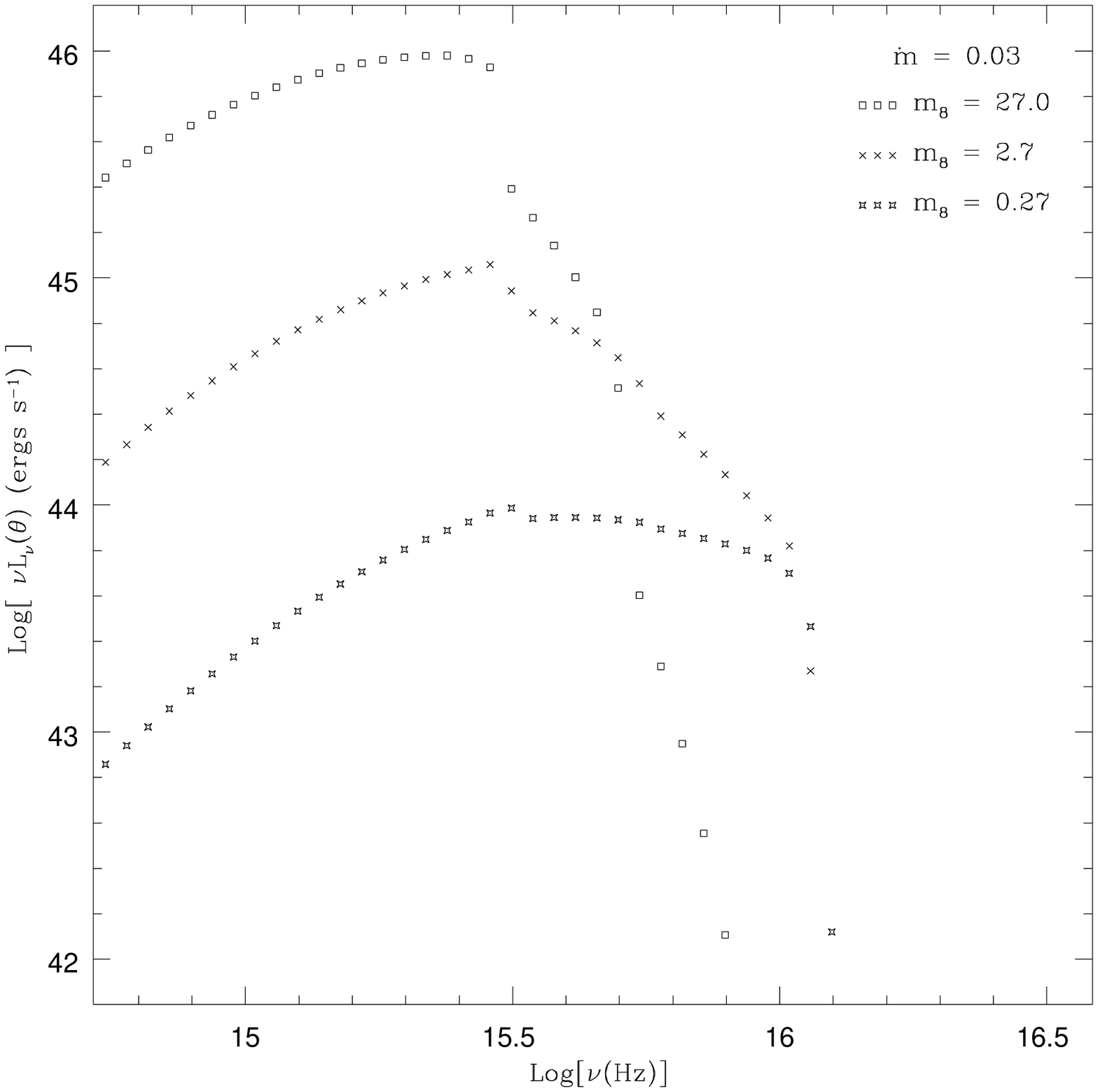}
\caption{\label{fig: standard mass} }
\end{figure}

\bigskip
Figure \ref{fig: standard mass}.  The face-on standard disk spectrum for 
a fixed
mass accretion rate and varying central mass.

\begin{figure}
\plotone{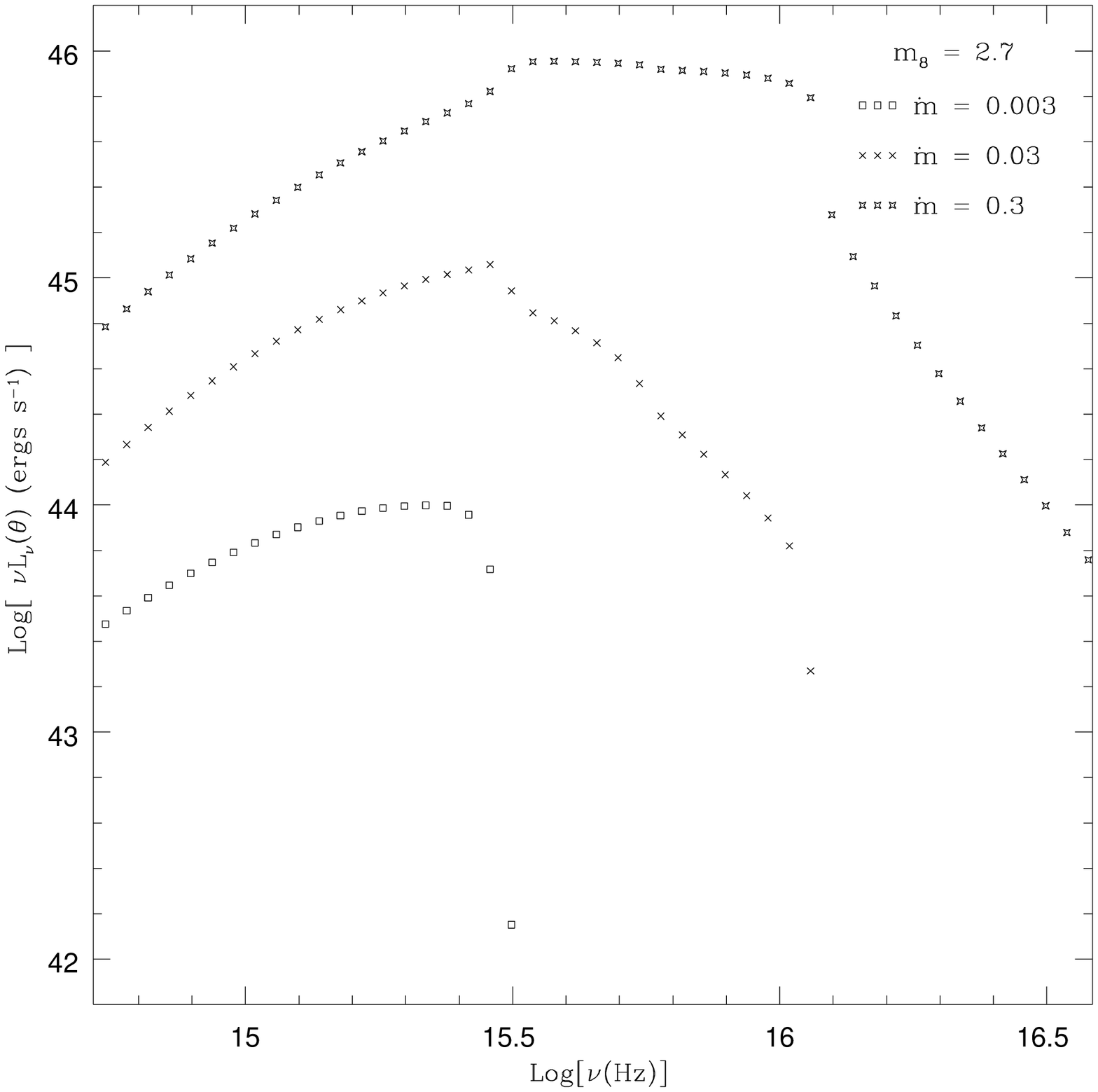}
\caption{\label{fig: standard mdot} }
\end{figure}

\bigskip
Figure \ref{fig: standard mdot}. The face-on standard disk spectrum 
for a fixed 
central mass and varying mass accretion rate.

\begin{figure}
\plotone{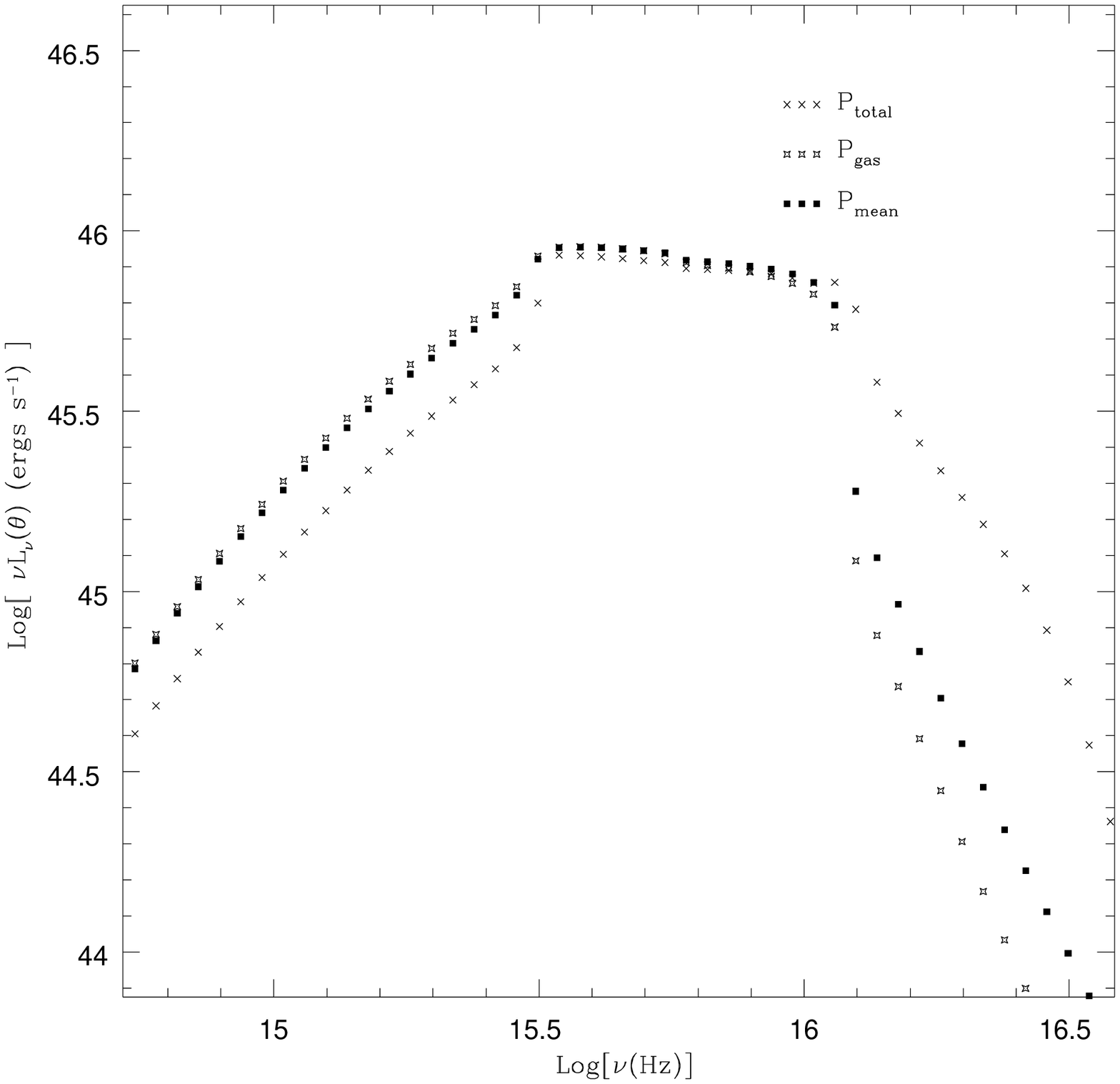}
\caption{\label{fig: standard visc} }
\end{figure}

\bigskip
Figure \ref{fig: standard visc}. The face-on standard disk spectrum as a
function of viscosity prescription.

\bigskip
Table \ref{table: spectral parameters}.  Parameters of the UV spectra.

\vfil\eject
\begin{table}
\begin{tabular}{||c|c|c|c|c|c||} \hline
$\dot m$ & $m_8$ & $\dot m/ m_8$ & $\alpha_n$ & $\alpha_{f}$ & $A_l$    \\
\hline
0.003     & 2.7     & $10^{-3}$  & -0.3       & $-\infty$    &  0.0     \\
0.03      & 0.27    & $10^{-1}$  &  0.4       &   -1.4       &  0.9     \\
0.03      & 2.7     & $10^{-2}$  &  0.1       &   -3.2       &  0.6     \\
0.03      & 27.0    & $10^{-3}$  & -0.6       &  $-\infty$   &  0.3     \\
0.3       & 2.7     & $10^{-1}$  &  0.2       &   -1.2       &  1.6 \\ \hline
\end{tabular}
\caption{\label{table: spectral parameters}}
\end{table}

\end{document}